\documentstyle[aps,preprint]{revtex}
\tightenlines

\newcommand{\be}{\begin{equation}}
\newcommand{\ee}{\end{equation}}
\newcommand{\bea}{\begin{eqnarray}}
\newcommand{\beas}{\begin{eqnarray*}}
\newcommand{\eea}{\end{eqnarray}}
\newcommand{\eeas}{\end{eqnarray*}} 
\newcommand{\ba}{\begin{array}}
\newcommand{\ea}{\end{array}}
\def\ls{\mathrel{\lower4pt\vbox{\lineskip=0pt\baselineskip=0pt
           \hbox{$<$}\hbox{$\sim$}}}}
\def\gs{\mathrel{\lower4pt\vbox{\lineskip=0pt\baselineskip=0pt
           \hbox{$>$}\hbox{$\sim$}}}}

\begin{document}

\draft

\preprint{}
   
\title{Baryogenesis in theories with large extra spatial dimensions}
\author{ Rouzbeh Allahverdi~$^{1}$, Kari Enqvist~$^{2}$, 
Anupam Mazumdar~$^{3}$ and
Abdel P\'erez-Lorenzana~$^{3,4}$}
\address{$^{1}$ Physik Department, TU Muenchen, James Frank
Strasse, D-85748, Garching, Germany. \\
$^{2}$ Department of Physics and Helsinki Institute of Physics,
P. O. Box 9, FIN-00014, University of Helsinki, Finland.\\
$^{3}$ The Abdus Salam International Centre for Theoretical Physics, I-34100,
Trieste, Italy.\\
$^{4}$ Departamento de F\'{\i}sica,
Centro de Investigaci\'on y de Estudios Avanzados del I.P.N.\\
Apdo. Post. 14-740, 07000, M\'exico, D.F., M\'exico.}


\maketitle

\begin{abstract}
We describe a simple and  predictive scenario for baryogenesis  in theories
with large extra dimensions which resembles Affleck-Dine baryogenesis. The
Affleck-Dine field is a complex  scalar field carrying a $U(1)_{\chi}$ charge
which is dynamically broken  after the end of inflation. This generates an
excess of $\chi$ over $\bar \chi$, which then decays into Standard Model
fermions to produce an excess of baryons over anti-baryons. Our model is very
constrained because the Affleck-Dine field has to be  sufficiently flat during
inflation. It is also a source for density  fluctuations which can be tested in
the coming satellite and balloon experiments.

\end{abstract}

\vskip90pt

\section{Introduction}

Recently it has been proposed that large extra spatial 
dimensions can explain the apparent weakness of the 
electroweak scale as compared to gravity in $3+1$ dimensions. 
In such a scenario four dimensional world is assumed to be a flat
hypersurface, called a  brane, which is  embedded in a higher dimensional
space-time, which is known as the bulk. The hierarchy 
problem is then resolved by assuming that  TeV scale can  be 
the fundamental scale in higher dimensions~\cite{nima0,early,abdel00}. 
This however requires the size of the extra dimensions to be large. 
Such a large volume can substantiate the hierarchy in energy scales. 
The volume suppression $V_{d}$, the effective four dimensional Planck mass 
$M_{\rm p}$, and, the fundamental scale in $4+d$ dimensions $M_{*}$ are all
related to each other by a simple mathematical relation 
\begin{equation}
M_{\rm p}^2 =M_{*}^{2+d}V_{d}\,.
\end{equation}
This automatically sets the present {\it common size} of all the extra
dimensions at $b_0$. For two extra dimensions, and, $M_{*}=1$ TeV, the required
size is of order $0.2$ mm right on the current experimental limit for the
search of deviations in Newton's gravity~\cite{exp2}.  Recent experimental
bounds suggest $M_\ast$ to be much larger. Naturally, such  model has an
important impact on collider experiments~\cite{exp1}, and on  cosmology. In
this paper we  address one of the most important issues in particle cosmology,
the origin of baryon asymmetry in theories with large extra dimensions.

\vskip7pt

The generation of baryon asymmetry requires three well-known conditions;  
$C$ and $CP$ violation, $B$ or $L$ violation, and out of equilibrium decay
\cite{sakh}.  It is quite probable that the early Universe  had strong 
departure from thermal equilibrium due to large expansion rate in the 
early Universe. However, achieving out of equilibrium condition
becomes more difficult when the inflationary scale and the 
final reheat temperature of the Universe is lowered down to the electroweak 
scale and below. As we shall see, this is a major obstacle for realizing
baryogenesis in the context of large extra dimensions. Another major
problem is a proton decay. A low fundamental scale induces fast proton 
decay via dimension $6$ baryon number violating operator in the Standard 
Model (SM). With a low fundamental scale the usual coupling suppression 
is not sufficient. 

\vskip7pt

The cosmological setup for large extra dimensions is quite different from the
conventional one. Firstly, if the electroweak scale is the fundamental  scale
in higher dimensions then there can be no massive fields beyond  the 
electroweak scale in four dimensions. Secondly, the size of the extra 
dimensions can be quite large as compared to the electroweak scale, which
implies a new degree of freedom with a small mass scale related to the size of
the extra  dimension. This field is usually known as the radion field. Its mass
can be as small as ${\cal O}(\rm eV)$ for two large extra dimensions. The
stabilization of such large  extra dimensions is a dynamical issue because they
grow from their natural  scale of compactification $\sim(\rm TeV)^{-1}$ to the
millimeter size  in order to solve the hierarchy problem. In fact, the
stabilization must take place  at the very initial phase of the Universe via
some trapping mechanism as discussed  in Ref.~\cite{abdel0}. 

\vskip7pt

Another challenge is how to realize inflation in these models such
that one naturally obtains the correct amplitude for the density
perturbations. There have been many proposals \cite{many}, but, the most
appealing one is invoking a SM singlet scalar living in the bulk
\cite{abdel1}, and we shall point out why this is the only mechanism which
works well. There is also the question of the presence of Kaluza Klein (KK)
states of the graviton and any other fields residing in the bulk.  At high
temperatures it is quite possible to excite these KK states. Above a certain
temperature known as the {\em normalcy temperature}, the Universe could be
filled by the KK modes. The Big Bang Nucleosynthesis (BBN) constrains  the
normalcy temperature to be above $\sim 1$ MeV. In order to be safe from other
cosmological bounds the final reheat temperature which is the  largest
temperature one can envisage during the radiation era, should be  smaller than
the normalcy temperature, which is constrained by cosmological  considerations
to be as small as $100$ MeV \cite{nima0,davidson0}.

\vskip7pt

There are many ways of generating baryon asymmetry in the Universe, one of
which is the simplest and predictive scheme known as the Affleck-Dine mechanism
\cite{affleck}, first discussed in the context of supersymmetry. A scalar
condensate which carries non-zero baryonic, or/and  leptonic charge survives
during inflation and decays into SM fermions to  provide a net baryon
asymmetry. The adaptation and elaboration of this particular mechanism in
theories with large extra dimensions is the main goal of this paper. It has
already been shown that the Affleck-Dine mechanism is the only
solution for providing an adequate baryon asymmetry in this context  with a
very low reheat temperature, see Ref.~\cite{abdel2}.

\vskip7pt

We begin our discussion by introducing some of the salient cosmological 
features of the large extra dimensions. In section II, we briefly discuss the 
consequences of having a temperature larger than the {\em normalcy
temperature}.  This is necessary in order to judge the merit of other
baryogenesis  scenarios such as leptogenesis and electroweak baryogenesis.  
In  section III we discuss inflation and its couplings to the SM fields. 
We shall also discuss very briefly the radion stabilization because of 
its importance. In section IV  we  consider various other proposals of 
baryogenesis and argue that within the context of large extra dimensions, 
where the inflaton is a higher dimensional  field, they essentially 
fail to provide the observed baryon asymmetry.  In section V we introduce 
our Affleck-Dine baryogenesis model which we embed within  an inflationary 
scenario. In section VI we describe various facets of our  model 
and argue why it appears to be the sole mechanism which can generate 
the correct baryon asymmetry. We shall also highlight various ways
of constraining the model parameters. Note even though in most of the
cases we specialize for the critical two extra dimensions, our
applications and conclusions are valid for any number of extra spatial
dimensions.  Finally, in section VII we summarize our main results.


\section{Thermal history of the Universe}

In order to appreciate thermal history of the Universe we need to estimate
the temperature below which the Universe could be safely regarded as a
radiation dominated one, which we call here the {\em normalcy temperature}, 
$T_{\rm c}$. In this  section we shall review some of the already known results
and provide some new insight on how the cosmological evolution changes if the
temperature of the Universe exceeds $T_{\rm c}$. It is possible to
excite the KK states from the plasma with a temperature $T$. The production of
these  states from relativistic particles depends on the cross section which
is Planck mass suppressed. It is noticeable that the cross section of  
processes such as $\gamma+\gamma \rightarrow G$ is of order 
$\sigma_{\gamma + \gamma \rightarrow G} \sim (TR)^{d}/M_{\rm p}^2$
\cite{nima0}, where $R$ is the effective size of the extra dimensions. Once
gravitons are  produced their evolution can be traced by the Boltzmann equation
\begin{equation}
\label{evol0}
\frac{\partial n_{G}}{\partial t} + 3H n_{G} = \langle \sigma~v\rangle
n^2_{\gamma} \sim \frac{T^{d+6}}{M_{*}^{d+2}}- \frac{n_{G,m}}{\tau_{G}}\,,
\end{equation}
where $\tau_{G}$ is the decay life time of a massive KK mode, given by
\begin{eqnarray}
\tau_{G} \approx \frac{M_{\rm p}^2}{m^3}\,.
\label{taum}
\end{eqnarray}
If we naively assume that right after the end of inflation the
first term is dominating the right-hand side of Eq.~(\ref{evol0}),
then by taking $a(t)T(t)={\it constant}$, where $a$ is the
scale factor of the Universe, we can in fact simplify the above equation.
While doing so we may also neglect the evolution
of the individual mode and shall concentrate upon all possible
KK states excited up to a given temperature. We obtain
\begin{eqnarray}
\label{evol1}
\frac{d(n_{G}/n_{\gamma})}{dT}=-\frac{\langle \sigma ~v\rangle n_{\gamma}}
{H ~T}\,.
\end{eqnarray}
Once the KK states are excited they are no longer in thermal equilibrium.
However, individual KK modes with a mass $\sim T$ remain present in the
Universe until they decay. We can integrate Eq.~(\ref{evol1}) while assuming
that we are in a standard cosmological era such that
$H^2\propto \rho/M_{\rm p}$, to get
\begin{eqnarray}
\label{final0}
\frac{n_{G}(T)}{n_{\gamma}}= \frac{n_{\gamma}(T_{\rm r})
\langle \sigma ~v\rangle}{H(T_{\rm r})}\,.
\end{eqnarray}
The temperature $T_{\rm r}$ designates the largest
temperature during  radiation era, known as the reheat
temperature of the Universe. We also take $v=1$, henceforth. Now substituting
the cross section and assuming that the relativistic particles
dominate the Universe; $n_{\gamma} \sim T_{\rm r}^3$,
we evaluate the right-hand side of Eq.~(\ref{final0}). The ratio thus
obtained can not exceed more than one at any later times in order
to maintain the successes of nucleosynthesis era and so we obtain
a simple bound on $T_{\rm r}$, which is given by \cite{nima0}
\begin{equation}
\label{final1}
T_{\rm r} \leq T_{\rm c} 
\sim \left(\frac{M_{*}^{d+2}}{M_{\rm p}}\right)^{1/1+d}\,.
\end{equation}
Note, for the preferred values  $M_{*} \sim 30$ TeV for $d=2$, we obtain
$T_{\rm c} \leq 100$ MeV. This is an extremely strong constraint on the
thermal  history of the Universe. It reiterates very strongly that the
radiation dominated  Universe simply can not prevail beyond this temperature.
Therefore, any physical phenomena such as first order phase transition,
out-of-equilibrium decay of heavy particles, if at all taking place beyond
$T_{\rm c}$,  need to be  revised. Now, it is pertinent to ask how the
evolution of the Universe changes if we assume that it is possible to exceed
the normalcy temperature, but  let us first note that as we take larger
number of extra dimensions, $d \rightarrow \infty$,  we apparently increase
$T_{\rm c}$ up to the fundamental scale $T_{\rm c} \rightarrow M_{*}$
\cite{nima0}.

\vskip7pt

A priori it can not be ruled out that the temperature of the Universe should 
not exceed $T_{\rm c}$. The reheat temperature is determined  by the inflaton
coupling to the matter fields. The only constraint upon $T_{\rm r}$ 
is that it must be larger than few MeV.  Therefore, there is no guarantee that
the Universe can not have a radiation domination with an instantaneous
temperature more than the normalcy temperature. In such a case, there is 
a plenty of KK states which can be excited easily beyond $T_{\rm c}$, which
actually leads to enhancing the number of relativistic degrees of freedom. The
number of relativistic degrees of freedom increases as $g(T) \sim (R T)^{d}$,
which also determines the number of degrees of freedom determining entropy. At
this point one may wonder how the KK modes, which actually have Planck mass 
suppressed couplings, can be brought into thermal equilibrium. Indeed, they
cannot if $n_{G} \ll n_{\gamma}$. However, we are in a opposite limit when the
KK gravitons have started dominating the number density of relativistic decay
products of inflaton. Once the KK modes are produced  they go out of
equilibrium due to the fact that the self interaction among these gravitons is
extremely weak. However, at the time they are being produced they introduce an
extra entropy to the already existing plasma. The entropy injection becomes
important once we are above $T_{\rm c}$ (the KK modes are excited  with masses
up to $T \gg T_{\rm c}$). The distribution function  for each and every KK mode
has a completely different profile, for the KK mode with mass 
$m \sim T_{\rm c}$ can be understood to be close to the relativistic 
particles, but the same can not be true for the heavier KK modes. As a 
result the effective entropy stored within the KK modes can be written as
\begin{eqnarray}
\label{entropy}
s =\frac{\rho+p}{T} \sim (RT)^{d}~T^3 \,,
\end{eqnarray}
neglecting factors ${\cal O}(1)$. The term in the bracket corresponds to the
relativistic degrees of freedom which is roughly the number of degrees of KK
modes. We should mention here that while deriving Eq.~(\ref{entropy}), we  have
naively assumed that the final distribution function for the KK modes are 
peaked around the final temperature and all the KK modes below that are 
produced abundantly with an uniform distribution which mimics that of 
a relativistic species. 

\vskip7pt

Now, following the fact that entropy conservation gives 
\begin{equation}
\label{cool0}
sa^3 = T^{d+3} a^3= {\it constant}\,,
\end{equation} 
we obtain a simple relationship 
between the expansion of the Universe and the rate of change of the temperature
\begin{eqnarray}
\label{cool1}
H=-\frac{d+3}{3}\frac{\dot T}{T}\,,
\end{eqnarray}
which actually determines the cooling rate of the Universe provided the KK
modes are fairly stable. This then leads to an approximate Hubble parameter
\begin{equation}
\label{hub00}
H(T) \approx \frac{T^{(d+4)/2} R^{d/2}}{M_{\rm p}}\,.
\end{equation}
However, the KK modes do decay and some of them would perhaps decay much
before nucleosynthesis, but in our case this depends on the maximum
temperature one could reach beyond the normalcy temperature. Note, in a
standard cosmology with a radiation dominated Universe the temperature scales
like inverse of the scale factor. In such a case any particle which has a
Planck mass suppressed couplings with other fields can decay before
nucleosynthesis provided its mass is beyond $10$ TeV. Otherwise, lighter
particles decay much after Nucleosynthesis and their number density is well
constrained from the diffusion of gamma rays in a micro wave background
radiation \cite{ratio}. However,  if  we had a KK dominated Universe, then
there might be a temperature-scale factor relationship which would be 
governed by Eq.~(\ref{cool1}).  This would also affect the life time of the 
massive KK modes. The simplest estimation for the masses of the KK modes 
which would decay before nucleosynthesis could be obtained by demanding 
that a KK mode with a mass just above the {\em normalcy temperature} should 
decay before nucleosynthesis. This determines the temperature of the 
Universe at the time of the decay. For instance, for $d=2$ extra 
dimensions we obtain
\begin{equation}
T_{\rm decay} \propto R^{-1}\,,
\end{equation}
where $R$ is the size of the extra dimensions. In order to be consistent, the 
above temperature ought to be more than $\sim {\cal O}(\rm MeV)$. This
restricts the size of the extra dimensions to be much smaller than  $\sim
10^{-10}$~mm. This translates into a lower bound for the fundamental scale
which is now increased to $M_{\ast}\gs 10^{8}$ GeV. This result is inconsistent
with our basic assumptions of having a low quantum gravity scale. This 
certainly may rule out the possibility of having large extra  
dimensions for $d=2$. The situation improves a little if there are 
more than two extra dimensions. For instance, if we take $d=6$ the 
above analysis gives a bound for the size $R\ls 2\times 10^{-14}$~mm, and, 
for the fundamental scale $M_*\gs 2\times 10^5$~GeV. Since, the normalcy
temperature also increases, which can be $\sim {\cal O}(1)$ TeV, baryogenesis 
may not be so troublesome provided we also take into account at least six 
extra dimensions. However, a simple hope like this seems to be a mirage 
once we realize that inflaton must reside in the bulk whose couplings 
automatically determine a reheat temperature below electro weak scale. 
This conclusion is actually quite robust and regardless the number of 
extra dimensions.

\vskip7pt

Our result asserts that we must reheat our Universe
below $T_{\rm c}$ given by Eq.~(\ref{final1}), in order not to excite the KK
modes with an over abundance.   We have noticed that  normalcy temperature
is the largest temperature of the Universe below which one can safely regard
the content of the Universe as a radiation dominated one.

\vskip30pt

\section{Inflation, reheating and thermalization}

\subsection{Inflation and density perturbations}

In order to provide a relatively small reheat temperature one automatically
requires very small couplings of the inflaton to the matter fields. As we 
already know that the dynamics of the inflaton plays a crucial role in
achieving out-of-equilibrium condition for baryogenesis.  Keeping this in
mind we briefly review inflationary dynamics which shall also act as a preview
for the AD baryogenesis.

\vskip7pt

As discussed above, the largest temperature in a radiation dominated Universe 
must be smaller than $100$ MeV for two large extra dimensions. One way to
achieve this is to assume that the inflaton is living in the higher
dimensional bulk, and upon compactification it naturally admits Planck
suppressed couplings to the SM fields. A simple model of inflation can be
constructed using only coupled  scalar fields in $4+d$ dimensions \cite{abdel1}
(for other attempts, see Ref.~\cite{many}). The  potential
can be written down as
 \begin{eqnarray}
  V(\hat N,\hat\phi) = {\lambda^2 M_*^d}
  \left(N_0^2 - {1\over M_*^d} \hat N^2\right)^2 +
  {m^2_{\phi}\over 2} \hat\phi^2 
  + {g^2\over M_*^d} \hat N^2 \hat\phi^2 \,,
 \label{hybp}
 \end{eqnarray}
where $\hat\phi$ is the inflaton field, and, $\hat N$ is the subsidiary field
which is responsible for the phase transition.  The coupling constants are $g$
and $\lambda$, in general they are different, and $N_0$ determines the
vacuum expectation value. Note that the higher dimensional field has a mass of
dimension $1+d/2$, which leads to non-renormalizable interaction terms.
However, the suppression is given by the fundamental scale instead of the four
dimensional Planck mass. Upon dimensional reduction the effective four
dimensional fields, $\phi, N$ are related to their higher dimensional relatives
by a simple scaling
 \be
\label{rel}
 \phi =  \sqrt{V_d} \hat\phi \,, \quad \quad N =\sqrt{V_d} \hat N\,.
 \ee
From the point of view of four dimensions the extra dimensions are assumed to
be compactified on a $d$ dimensional Ricci flat manifold with a radii $b(t)$,
which have a minimum at $b_0$. The higher dimensional metric then reads
 \begin{eqnarray}
 ds^2 = dt^2 -a^2(t)  d\vec x^2 -b^2(t)d\vec y^2\,,
 \end{eqnarray}
where $\vec x$ denotes three spatial dimensions, and $\vec y$
collectively denote the extra dimensions. The scale factor of a four
dimensional space-time is denoted by $a(t)$. After dimensional
reduction the effective four dimensional action reads 
 \begin{eqnarray}
 \label{action}
 S=\int d^4x \sqrt{-g}\left[ -\frac{M_{\rm p}^2}{16\pi}R + \frac{1}{2}
 \partial_{\mu} \sigma \partial^{\mu}\sigma -U(\sigma)
 +\frac{1}{2}\partial_\mu \phi\partial^\mu \phi +
 \frac{1}{2}\partial_\mu N\partial^\mu N \right. \nonumber \\ 
\left. -\exp(-d\sigma/\sigma_0)V(\phi, N) \right]\,.
 \end{eqnarray}
The potential $V(\phi, N)$ can be derived from Eqs.~(\ref{hybp}) 
and (\ref{rel}). 
\begin{equation}
\label{hybp0}
V(\phi, N) \equiv \left(\frac{M_{\rm p}}{M_{*}}\right)^2\lambda^2 N_0^4
+\frac{\lambda^2}{4}\left(\frac{M_{\ast}}{M_{\rm p}}\right)^2N^4-
\lambda^2 N_0^2N^2
+g^2\left(\frac{M_{\ast}}{M_{\rm p}}\right)^2\phi^2N^2 +
\frac{1}{2}m_{\phi}^2\phi^2\,.
\end{equation}
The radion field  $\sigma (t)$ can be written in terms of the radii of 
the extra dimensions
\begin{eqnarray}
\label{radion0}
\sigma(t) = \sigma_0 \ln \left[ \frac{b(t)}{b_0}\right]\,, \quad
\sigma_0 =
\left[\frac{d(d+2)M_{\rm p}^2}{16 \pi}\right]^{1/2} \,.
\end{eqnarray}
From the above equation, it is evident that $\sigma_0$ is proportional to the
four dimensional Planck mass. For $b(t) \sim ({\rm TeV})^{-1}$, and, 
$b_0 \sim 1$mm, the modulus of the radion field takes a very large initial 
value. The radion field has a potential which at the minimum is   
given by $U(\sigma)\sim m_{r}^2 \sigma^2$, where 
$m_{\rm r} \sim 10^{-2}{\rm eV}$. In this paper we
shall assume that such a potential is essential to stabilize the large extra
dimensions (for discussion, see Ref.~\cite{albrecht}).

\vskip7pt

There are mainly two phases of subsequent inflation in this scenario.
We notice that initially $\hat N$ is stuck in the false vacuum determined by
\begin{equation}
\label{hybp1}
V \approx \left(\frac{M_{\rm p}}{M_{\ast}}\right)^2\lambda^2 N_0^4+\frac{1}{2}
m_{\phi}^2\phi^2\,,
\end{equation} 
which renders $V(\phi,N) \approx \lambda^2 M_{\rm p}^2 M_{\ast}^2$ if 
$N_0 \sim M_{\ast}$, the contribution is almost constant if the vacuum 
dominates over the second term in Eq.~(\ref{hybp1}), which is necessary 
for the generation of density perturbations. The first phase of inflation 
occurs when $\phi$ is still rolling down the potential and has not undergone
a second order phase transition. During the first phase of inflation
the exponential term due to the radion field present in front of 
the scalar potential, see Eq.~(\ref{action}) is responsible for driving a
power law inflation. For details we refer \cite{abdel0}.  
The exponential potential inflates not only the brane but also the bulk,
which expands from $(\rm TeV)^{-1}$ to mm size, which is equivalent to
$35$ e-foldings of inflation if there are only two extra dimensions. 
Towards the end of this phase when the radion field 
$\sigma \rightarrow \sigma_0$, the radion gets a  
running mass $m_{\rm eff}^2 \approx {\cal O}(1) H^2 + m_{r}^2$.
The Hubble parameter $H^2 \gg m_{\rm r}^2$, therefore the Hubble correction 
dominates the radion mass. Once, the radion gains this mass it 
simply rolls down to its minimum as dictated by $U(\sigma)$. This ensures 
that not only the extra dimensions have already been grown to the 
adequate size $b_{0} \sim $mm, but also have been stabilized. This 
takes place while the inflaton $\hat \phi$ is still rolling down the 
potential and $\hat N$  is still locked in the false vacuum. From
the last term in Eq.~(\ref{action}) one finds that the exponential 
$\exp(\sigma/\sigma_0)$ becomes of ${\cal O}(1)$ when 
$\sigma$ is trapped in its minimum, and the effective potential 
is solely given by Eq.~(\ref{hybp1}).
Therefore the dynamics of $\sigma$ is completely frozen except for 
quantum fluctuations. 

\vskip7pt

In a completely model independent way it can be easily argued that
at least $43$ e-foldings of inflation is necessary in order to produce
structure formation \cite{abdel0}. The amount of
required inflation is less than the usual $60$ e-folding because of a
smaller inflationary scale and reheat temperature.
The scale of inflation during this second phase is simply given by 
\begin{equation}
\label{hub0}
H_0 \approx \sqrt{\frac{8\pi}{3}}\frac{\lambda N_0^2}{M_{\ast}}\,,
\end{equation}
If $N_{0} \sim M_{\ast}$, we automatically get the Hubble expansion at the end 
of inflation: $H_0 \approx M_{\ast}$. The adiabatic density perturbations can
be generated while the inflaton $\phi$ is rolling down the potential until it
reaches a critical value where a second order phase transition takes place and
the effective mass square for $N$ field becomes negative. The critical point
can be found from the potential Eq.~(\ref{hybp1}) to be
\begin{equation}
\label{critical} 
\phi_{\rm c} = \frac{\lambda}{g}\left(\frac{N_0 M_{\rm p}}{M_{\ast}}\right)\,.
\end{equation}
Note, if 
$\lambda \sim g $, and, $N_0 \sim M_{\ast}$, we automatically get 
$\phi_{\rm c} \sim M_{\rm p}$. 

\vskip7pt

The two slow-roll conditions for inflation can be laid down very easily
\begin{eqnarray}
\label{slowroll1}
\epsilon = \frac{M_{\rm p}^2}{16\pi}
 \left(\frac{V^{\prime}(\phi, N)}{V(\phi, N)}\right)^2
=\frac{1}{16\pi}\left(\frac{M_{\ast}^2 \phi}{M_{\rm p} N_0^2}\right)^2
\left(\frac{m_{\phi}}{\lambda N_0}\right)^4 \ll 1\,, \\
\label{slowroll2}
|\eta| = \frac{M_{\rm p}^2}{8\pi}\left \arrowvert 
 \frac{V^{\prime \prime}(\phi, N)}
 {V(\phi, N)}\right \arrowvert = \frac{1}{8\pi}
 \left(\frac{M_{\ast}}{N_0}\right)^2
 \left(\frac{m_{\phi}}{\lambda N_0}\right)^2 \ll 1 \,,
\end{eqnarray}
where prime denotes derivative with respect to $\phi$. The spectrum of the 
density perturbations is given by \cite{liddle}
\begin{eqnarray}
\delta_{\rm H}^2 = \frac{32}{75}\frac{V(\phi, N)}{M_{\rm p}^4} \frac{1}
{\epsilon_{43}}\,,
\end{eqnarray}
where $\epsilon_{43}$ is defined roughly $43$  e-foldings before the end 
of second phase of inflation. From data 
$\delta_{\rm H} \simeq 1.91 \times 10^{-5}$ \cite{bunn}. The 
interesting cosmological scale leaves the horizon when 
$\phi_{43}= \phi_{\rm c}e^{43|\eta|} \approx \phi_{\rm c}$, 
provided we assume $|\eta| \ll 1/43$. By proper substitution we get 
a simple relationship 
\begin{equation}
\delta_{\rm H} =8.2\lambda^2 g \frac{N_0^5}{M_{\ast}^2 m_{\phi}^2 M_{\rm p}}\,.
\end{equation}
In order to illustrate, let us set $N_0 = M_{\ast} \sim 100$ TeV and 
$m_{\phi}\sim 10$ GeV; we then obtain 
\begin{equation}
\delta_{\rm H} \sim (\lambda^2 g )\times 10^{-5}\,,
\end{equation}
which  provides the right COBE (and Boomerang)
amplitude with $\lambda \sim g \sim {\cal O}(1)$. Let us stress that all 
the above conclusions are independent of the number of extra dimensions.

\vskip7pt

Let us mention here that we actually require  $N_{0}<M_{\ast}$ for two
reasons. First, $N$ provides a mass squared contribution to the Higgs
potential, which has to be less than  the electro-weak scale in order to
provide the right magnitude for the Higgs mass. The second obvious point is
that effective mass for $\phi$ and $N$ must be greater than  the Hubble rate
$H$ after the end of inflation. This is necessary to terminate inflation just
after the phase transition. However, for most of our calculation  $N_0 \sim
M_{\ast}$ remains a very good approximation in order to show the merit of the
baryogenesis model.

\vskip7pt

The spectral index $n= 1+2\eta -6\epsilon$ is presently constrained by
observations  $|n-1|< 0.13$. If we naively assume $N_0 \equiv M_{\ast}$, and,
$\lambda \equiv g$, 
we find
\begin{equation}
\label{limit}
\lambda M_{\ast} \sim 1.27 \times 10^{15}|\eta|~ {\rm GeV}\,.
\end{equation}
This suggests that $|\eta|$ has to be extremely small. Given that, $\epsilon$
is also  very small, see Eq.~(\ref{slowroll1}), we conclude that our model
predicts a  perfect scale invariant density perturbations. If the coupling is
of order one and $M_{\ast} \sim 100$ TeV, the slow roll parameter 
\begin{equation}
\label{con}
|\eta| \leq 10^{-10}\,.
\end{equation} 
If we assume that the inflaton sector is solely responsible
for the adiabatic density perturbations, the small slow roll parameter 
constrains the amplitude of any other scalar fields we intend to 
introduce in our setup. This we shall elaborate when we discuss AD
baryogenesis.


\subsection{Post inflationary dynamics of $\phi$ and $N$}

Let us first discuss the classical dynamics of the fields $\phi$ and $N$
after the phase transition. During this era both the fields oscillate
with an initial amplitude determined by
\begin{equation}
\label{amp0}
N =\sqrt{2}\left(\frac{M_{\rm p}}{M_{\ast}}\right)N_0\,, 
\quad \quad \quad \phi=
\phi_{\rm c}=\frac{\lambda}{g}\left(\frac{M_{\rm p}}{M_{\ast}}\right)N_0\,.
\end{equation}
The frequencies are determined by the effective mass scales
of the fields at the global minimum, given by
\begin{equation}
\label{amp1}
\bar m_{\phi} = \sqrt{2}gN_0 \,, \quad  \bar m_{N}=\sqrt{2}\lambda N_0\,.
\end{equation}
We notice that the phase transition must terminate inflation 
immediately, otherwise there could be another bout of inflation which might
give rise a particular signature in the spectrum of the microwave
background radiation which we will discuss in a separate publication. 
A slow transition might provide density perturbations of order one and 
might produce primordial black hole formation. In this paper we do not 
take into account of this phase. Therefore, in order to ensure 
$\bar m_{\phi} \sim \bar m_{N} > H_0$, the constraint on 
$N_0 < \sqrt{6/8\pi}M_{\ast}$, for $\lambda \sim g$. 
However, in order to illustrate our model for baryogenesis the essential
physics remains for $N_0 \approx M_{\ast}$.

\vskip7pt

When the frequencies of the oscillations are different the fields loose their 
coherence and the motion becomes chaotic, see Ref.~\cite{garcia,mar0,dan}. If
the coupling strengths $g$ and $\lambda$ are equal we
can easily obtain the field trajectories around the bottom of the potential,
which follows a straight line in a phase space of $\phi -N$ \cite{mar0,dan}.
There exists a particular  solution of the classical equations of motion given
by \cite{mar0,dan}
\begin{equation}
\label{ev0}
N(t)=\sqrt{2}(\phi_c-\phi(t))~\,.
\end{equation}
It can be argued that in a static Universe it is possible to obtain
a classical solution for either $\phi$ or $N$ near the bottom of the 
potential. Initial motion is an-harmonic, but the 
oscillations near $N_0$ can be expressed in a simple form:
\begin{equation}
\label{ev1}
N(t) \approx \sqrt{2}\left(\frac{M_{\rm p}}{M_{\ast}}\right)N_0 \left[1+ 
\frac{1}{3}\cos(m_{\phi}~t)\right]\,.
\end{equation}
In an expanding Universe the amplitudes of the oscillations decay and
\begin{equation}
\label{ev2}
N(t) \approx \sqrt{2}\left(\frac{M_{\rm p}}{M_{\ast}}\right)N_0 \left[ 1+ 
\frac{\Phi(t)}{3}\cos(m_{\phi}~t)\right]\,,
\end{equation}
where the amplitude of the oscillations decreases as $\Phi(t)\propto 1/t$. The
precise form of $\Phi(t)$ depends on the ratio $H/\bar m_{\phi}$. 
If the ratio is
large, the decay of the amplitude is felt in a  couple of oscillations,
otherwise, it may take many oscillations before the  expansion leads to
decaying in amplitude. In our case we can make a rough estimation.  We notice
that $\bar m_{\phi} \sim \bar m_{N} \sim H_0$ at the time of phase transition. 
Therefore, the fields begin to roll down towards their respective   minimum of
the potential from their initial amplitude which is  $\sim M_{\rm p}$. The 
amplitude of the fields decreases very quickly until  it reaches a point when
$\bar m_{\phi} \sim \bar m_{N} \geq H$. This happens  because the Hubble
parameter $H$ is also decreasing as the fields roll down  as $\sim 1/t$, where
$t$ is the physical time starting from the end of inflation.  In order to see
this, we need to ensure 
\begin{equation}
\frac{H(t)}{H_0} \equiv \frac{1}{tH_0} \ll 1\,,
\end{equation}
which happens when 
\begin{equation}
\frac{1}{H_0} \ll t \equiv \frac{2\pi N_{\rm osc}}{\bar m_{\phi}}\,,
\end{equation}
where $N_{\rm osc}$ is an approximate number of oscillations.  For $\bar
m_{\phi} \sim H_0$, we obtain  $N_{\rm osc} > 1/2\pi$. Therefore, within one
oscillation the Hubble parameter decreases quite rapidly. This  makes sure that
the fields having masses $\bar m_{\phi} \sim \bar m_{N} \sim H_0$  can
oscillate about their minimum with a common frequency for many oscillations
before they completely decay. We may estimate the number of oscillations to be
$N_{\rm osc} \sim \bar m_{\phi} \tau \equiv (M_{\rm p}^2/M_{\ast}^2) \sim
10^{28}$. Hence during the initial stages the dominant effect of expansion 
will be to render the total oscillations more harmonic around the minimum.
Therefore, our assumption of the field evolution given by Eq.~(\ref{ev2}) holds
well. However, in reality we need the opposite limit on masses, which would
only slow down the decay of the amplitude of the oscillations. With this brief
discussion on  the dynamics of the fields we move on to discussing
thermalization.


\subsection{Reheating and thermalization}

It is believed that the total energy density of the inflaton is  transferred
into  radiation. The minimal requirement is to have a thermal bath with a
temperature more than ${\cal O}(1)$ MeV in order to preserve the successes  of
BBN. Recall that now we can not reheat the Universe above normalcy temperature
$T_{\rm c}$. The final reheat temperature depends on the decay rates
$\Gamma_{\phi,N}$ of the oscillating field \cite{abdel1} 
\begin{equation}
T_{\rm r} \sim 0.1 \sqrt{(\Gamma_{\phi}+\Gamma_{N})M_{\rm p}}\,.
\end{equation} 
Therefore, we need extremely weak, non-renormalizable  coupling of $\phi$ and
$N$ to other SM fields. In our model this is natural because the  bulk fields 
have Planck suppressed interactions with matter fields stuck on the brane.
However, this fact also causes some problems, e.g. why the zero mode inflaton
is reheating the brane and why not the bulk? The inflaton could decay into some
other lighter degrees of freedom into the bulk. This point has been already
addressed in Ref.~\cite{abdel1} for the case of gravitons, and here we 
recapitulate the arguments.

\vskip7pt

In our previous section we found that both the fields get an effective mass 
term $\bar m_{\phi} \sim \bar m_{N} \propto \sqrt{2}\lambda N_0$, if we assume
$\lambda = g$. Setting $N_0 \sim {\cal O}(\rm TeV)$, then both $\phi$ and $N$
are kinematically allowed to decay into Higgs field, $h$, with a 
mass ${\cal O}(100 \rm GeV)$. 
This is different from the usual Kaluza-Klein theories
where the production of matter through inflaton decay occurs everywhere.
The decay rate is estimated as follows~\cite{abdel1}
\begin{equation}
\label{decay0}
\Gamma_{\phi, N \rightarrow hh}
	\sim \frac{f^2 M_{\ast}^4}{32\pi M_{\rm p}^2 m_{\phi,N}}\,,
\end{equation}
where, $f$ is the coupling constant that we take of order one. 
If $\bar m_{\phi}\sim \bar m_{N} \approx 0.1M_{\ast}$, 
then the reheat temperature is $T_{\rm r}\leq 100$~MeV, which is more than
${\cal O}(1)$ MeV and below the normalcy temperature, even for 
the most constrained case of $d=2$ extra dimensions. However, note
that such a low reheat temperatures is a generic prediction if the inflaton
field is living in the bulk.

\vskip7pt

The next point is concerning the production of KK graviton via the inflaton
decay.  The KK gravitons can be directly produced from the KK modes of 
$\phi$ and $N$, and possibly the KK modes of other scalar fields present 
in the bulk via  $\phi_{n} \rightarrow \phi_{l}G_{n-l}$
interactions, where $n,l$ are the KK numbers, and $G$ is the KK graviton. On
the other hand $\phi_{n}$ modes can be produced via collision processes, such
as $\phi \phi \rightarrow \phi_{n} \phi_{-n}$, and  similar reaction for $N$.
The rate of exciting the KK modes and their subsequent decay rate to graviton
can be estimated as \cite{abdel1}
\begin{equation}
\label{decay1}
\sigma_{\phi\phi \rightarrow \phi_{n}\phi_{-n}}\sim \lambda^2 \frac{M_{\ast}^2}
{M_{\rm p}^4}\,, \quad \quad \Gamma_{\phi_{n}\rightarrow \phi_{l}G_{n-l}}
\sim \frac{m_{n}m_{l}^2}{12\pi M_{\rm p}^2}\,,
\end{equation}
where $m_{n}^2=\bar m_{\phi}^2+n^2/R^2$ is the excited KK mode. Note that the
$NN$ or $\phi\phi$ scattering rates for producing their KK counterparts are
smaller than the direct decay of $\phi,~N$ to the brane fields. 
Therefore, the  zero modes of $\phi$ and $N$ still prefers the Higgs as 
a final decay product. However, late during the thermalization era 
the scattering phenomena may give rise to the production of KK modes 
of inflaton. In the above equation we have only estimated a single 
process for graviton production. There are plenty of other 
accessible modes in the final channel. This enhances the decay rate
\begin{equation}
\Gamma_{\phi_{n},~{\rm total}}=
\sum_{l}\Gamma_{\phi_{n}\rightarrow \phi_{l}G_{n-l}}
\sim \frac{m_{n}^3}{12\pi M_{\ast}^2}\,.
\end{equation}
The excited KK inflaton or KK partners of other scalar fields
present  in the spectrum are extremely short lived. The heavier KK mode decays
into the lighter KK mode plus gravitons, and eventually all the KK modes of
the inflaton decay into the zero mode. In case where extra bulk fields 
are present the reheating of the bulk can be naturally avoided, if 
either those modes are as heavy as the inflaton, or, the effective 
inflaton couplings to the bulk fields are smaller than the 
inflaton-brane interactions. 

\vskip7pt

So far we have tacitly assumed that reheating is almost instantaneous. This 
might not be the case, especially when the fields are oscillating. During this
period, the equation of state of the Universe in most of the cases is given by
that of a matter dominated era. However, if $\phi,N$ are decaying very slowly,
then they might also decay into lighter degrees of freedom, such as
relativistic species directly. One might expect that this could be the most
preferred channel. However, this is not the case, because again the oscillating
fields have non-renormalizable couplings to these lighter fields and the
decay rate follows
$\Gamma_{\phi,N \rightarrow \gamma \gamma}\sim M_{\ast}^3/M_{\rm p}^2$. 
Notice, its resemblance with that of Eq.~(\ref{decay0}),  
if $\bar m_{\phi} \sim M_{\ast}$. This is an important lesson and all it tells
us that $\phi,N$ decaying into Higgses and into lighter degrees of freedom is
equally preferred. However, if there were some radiation which could
thermalize, then in such a case, the Universe could in principle follow an
equation of state which would be determined mainly by a mixture of relativistic
species, inflaton, and non-relativistic Higgses. This would only affect the
thermal history of the Universe, and not the dynamics, because inflaton energy
density is dominating over all. It has been shown in the simplest situation
where inflaton and radiation components are allowed, that the instantaneous
temperature of a plasma might exceed the reheat temperature of the Universe. In
order this to happen one must also satisfy $H \geq \Gamma_{\phi}+\Gamma_{N}$
\cite{ratio}. We notice that this condition may be satisfied in our case at the
very beginning of the oscillations, but in spite of large initial amplitudes for
$\phi,N$, the oscillations are damped quickly.

\vskip7pt

The temperature of the plasma may reach its maximum
when $a/a_{0} \sim 1.48$, where $a$ denotes the scale factor
of the Universe and the subscript ${0}$ denotes the era when inflation comes
to an end. The maximum temperature is given by \cite{chung} 
\begin{equation}
\label{max0}
\frac{T_{\rm max}}{T_{\rm r}} =0.77 \left(\frac{9}{5\pi^3g_{\ast}}\right)^{1/8}
\left(\frac{H_0 M_{\rm p}}{T^2_{\rm r}}\right)^{1/4}\,,
\end{equation}
where $g_{\ast}$ denotes the effective number of relativistic degrees of
freedom.  For the purpose of illustration if we fix $H_{0} \sim M_{\ast}$ and 
$T_{\rm r} \sim 100$~MeV we found
$T_{\rm max} \sim 10^{5}$ GeV. This temperature
is much higher than the actual reheat temperature, and it seems that 
this is a generic prediction of inflationary scenarios in
extra dimensions. However, this  warrants
a preferred production of relativistic species from the decay of inflaton. 
This is unfortunately certainly not the case with the present situation. 
In order to proceed with our 
present discussion, we note that after reaching the maximum temperature, the 
temperature of the intermediate plasma decreases as
\begin{equation}
\label{max1}
T \sim 1.3 \left(
  \frac{g_{\ast}(T_{\rm max})}{g_{\ast}(T)}\right)^{1/4}T_{\rm max}a^{-3/8}\,,
\end{equation}
The important thing to notice here is that the thermal history of the Universe
is again different:  the temperature does not drop like $T\propto a^{-1}$ so
that the entropy of the plasma $s \propto a^{15/8}$. This will eventually
dilute any KK graviton being produced during this era, and,  as long as
$n_{G}/n_{\gamma} < 1$, we would not 
expect any further alteration of the results mentioned in Eqs.~(\ref{max0})
and (\ref{max1}).


\section{Baryogenesis}

The constraints on inflationary parameters which we have been discussing so
far must be bear in mind when discussing baryogenesis. We certainly need a
concrete mechanism from particle physics in order to address baryogenesis. 
The problem here is that we should introduce some 
baryon number violating processes. However, in theories where the 
fundamental scale is low naturally introduces dangerous 
higher order operators.
For instance,  dimension 6 operators such as $QQQL/\Lambda^2$,  
where $Q$'s and $L$ correspond quark and lepton SM doublets, which
can mediate  proton decay,  unless $\Lambda \geq 10^{15}$~GeV. 
This operator violates baryon and lepton number conservation by  
$\Delta B =\Delta L =1$. There is also a possibility of having 
right handed singlets $uuude/\Lambda^2$, for which $\Lambda \geq 10^{12}$ GeV.
There are other processes which violate baryon number, such as
neutron anti-neutron oscillations with a dimension $7$
operator  $uddudd/\Lambda^3$, which implies $\Lambda \geq 10^{5}$ GeV.
An alternative is the dimension $9$ operator 
$QQHQQH/\Lambda^5$ with $\Lambda \geq 10^{4}$. (The experimental bound 
on proton life time is $\tau_{p} \geq 10^{33}$ years~\cite{kam}, and for 
neutron anti-neutron oscillations $\tau_{n\bar n} > 1.2 \times 10^{8}$ 
seconds~\cite{nnbar,nnbar2}.)

\vskip7pt

We assume that the above mentioned operators can be avoided in some way
or other. It is very difficult to come up with a model where baryon 
violating operators are not constrained at a TeV scale. Thus, one has 
to ensure that such operators are not being reintroduced by the mechanism of 
baryogenesis. Especially, in our case we ought to be
careful with an  operator such as $\chi QQQL$. 
In case $\chi$ develops a vacuum expectation value $\sim M_{\ast}$, fast 
proton decay is inevitable.

\vskip7pt

Moreover, in the context of extra dimensions, leptogenesis is not a viable
mechanism  as we shall argue now. In leptogenesis a net lepton number is
produced in the decay  process of a heavy fermionic singlet such as a  right
handed neutrino, which is then  processed into baryon number by anomalous 
$B+L$ violating sphaleron interactions  \cite{tooft}. However, the electro
weak sphaleron transitions are active only  up to $100$ GeV \cite{kleb}. In our
case viability of leptogenesis has a simple catch. A singlet right handed
neutrino can naturally couple to the SM lepton doublet  and the Higgs field
through $y \bar{L} HN$~. This leads to a potentially large  Dirac mass term
unless the Yukawa coupling $y\sim 10^{-12}$. Moreover,  now the  see-saw
mechanism fails to work, since, the largest  Majorana mass we may expect  can
never be larger than  the fundamental scale. Therefore, given a very small
neutrino mass  $\sim y^2 \langle H\rangle^2/M \sim y^2\cdot {\cal O}(1)$ GeV, 
we still have to fine tune $y^2 \ls 10^{-10}$, in order to obtain the right  
order of magnitude for the neutrino mass. In any case the decay rate of the 
right handed neutrino is very suppressed and it is similar to
Eq.~(\ref{decay0}).  This means that when the right handed neutrino decays into
the SM fields,  the background temperature is of order of the reheat
temperature  $\sim {\cal O}(1-10)$~MeV. At this temperature the sphaleron rate
is exponentially suppressed, which is actually a set back for leptogenesis. A 
reheat temperature of at least ${\cal O}(1-100)$~GeV is required for making 
this scenario viable~\cite{pilaftsis}, which could in principle be attained if 
the number of extra spatial dimensions is increased to six, at least from the
point of view of the normalcy temperature. However, as we observed before, in
the class of models we considering this is not the case. Indeed, as mentioned
above already,  unless we increase the fundamental scale, the largest reheating
temperature we can  get from the inflaton decay is just barely  about 100 MeV, 
regardless the number of extra dimensions. This  makes leptogenesis even more
difficult.

\vskip7pt

A different possibility that sphalerons can reprocess a
pre-existing charge asymmetry into baryon asymmetry~\cite{olive}
reflected in an excess of $e_L$ over anti-$e_R$ created during
inflaton oscillations. This mechanism requires $(B+L)$ violating
processes to be out of equilibrium before $e_R$ comes into chemical
equilibrium, such that the created baryon asymmetry could be preserved.
Again, this has to happen at or above $100$ GeV. One could 
then assume that inflaton decays preferably into relativistic species
such that the plasma thermalizes to a temperature above $100$ GeV
\cite{davidson1,joyce}. However, this scenario cannot be implemented  
in the context of large extra dimensions because in this case the 
oscillating inflaton field injects more entropy to the thermal bath 
as discussed earlier. This leads to an immediate dilution of whatever 
baryon asymmetry has been created prior to the reheating era.  
The dilution factor is given by :
\be
\label{dilute}
\gamma^{-1}=\left({s(T_{\rm r})\over s(T_{\rm EW})}\right)
 =\left({g_{*}(T_{\rm r})\over g_{*}(T_{\rm EW})}\right)
 \left({T_{\rm r}\over T_{\rm EW}}\right)^3
 \left({a(T_{\rm r})\over a(T_{\rm EW})}\right)^3~,
 \ee
where $s$ is the entropy and $T_{\rm EW}\sim 100$ GeV. For a lower reheat 
temperature such as $T_{\rm r} \sim 1$ MeV, the above expression  gives 
$\gamma^{-1} \gs 10^{25}$ (assuming $T\propto a^{-3/8}$ and $g_{*}(T_{\rm c})
\approx g_{*}(T_{\rm r})$). Therefore, one concludes that the initial 
$n_{b}/s$ has to be extremely large $\gs 10^{15}$, this is certainly an
extraordinary requirement on any natural model of baryogenesis and practically 
impossible to achieve. Even if we increase reheat temperature
$T_{\rm r} \sim 100$ MeV, we would still require $n_{b}/s$ of order one.

\vskip7pt

We have learned two important lessons. First, the large entropy production 
during the last stages of reheating can in principle wash away any baryon 
asymmetry produced before electroweak scale. Second, it is extremely 
unlikely that leptogenesis would work. The only simple choice left is 
to produce  baryon asymmetry directly.  The sole mechanism which can 
be successful in these circumstances seems to be the Affleck-Dine (AD) 
baryogenesis\cite{affleck}, which we are going to discuss now.


\section{Affleck-Dine baryogenesis}

For the details of the AD baryogenesis we refer the readers to 
Ref.~\cite{lisa}. In our case the relevant questions are following:

\vskip7pt

$(1)$ {\it Can we have a condensate ?}

Unlike in the usual AD mechanism, which is  based on Minimal Supersymmetric 
Standard Model, we do not have flat directions automatically in-built  in our
model, or, protected by supersymmetry. Even if we invoke some  flat directions,
it cannot be associated with a condensate carrying $B$ or $L$.  The condensate
can not be protected alone due to lack of any symmetry argument.  One can not
form a SM condensate in the bulk because bulk gravity is color and  flavor
blind. This suggests that we necessarily have to assume some fundamental scalar
field.

\vskip7pt

$(2)$ {\it  What charge should it carry ?}

The AD field, which we denote here by $\chi$ has to be a gauge singlet
carrying  some global charge under $U(1)_{\chi}$. This global charge has to be
broken  dynamically in order to provide a small asymmetry in the current
density.  This shall be reflected by generating an excess of $\chi$ over 
$\bar \chi$. The charge associated with this AD field must be such that  baryon
number is violated maximally.  All that we need is to ensure that the SM quarks
maintain the small asymmetry between baryons and anti baryons. Notice that
$\chi$ field as such does not create the baryon asymmetry. The asymmetry is
produced due to the difference in number density of $\chi$ over $\bar \chi$.
This is the most important aspect of our model. This small asymmetry is then
transferred via the decay of $\chi$ and $\bar \chi$ into the SM quarks and
leptons. This constraints  the decay channel for $\chi$ and $\bar \chi$ which
we discuss next.

\vskip7pt

$(3)$ {\it What kind of interactions should it have ?}

This is a non-trivial issue because we do not have a condensate made up of 
SM quarks and leptons. We need to assume that $\chi$ interactions with SM 
fields conserve $U(1)_\chi$ symmetry. Therefore, the quarks and leptons must 
carry a non zero global $\chi$ charge. However, we do not want the 
$\chi$-$\bar\chi$ asymmetry to be transferred to non-baryon number violating 
interactions such as interactions involving the Higgses. Therefore,
the Higgs field should not carry a global $\chi$ charge, forbidding
$\chi$ decay into Higgses. 

\vskip7pt

Regarding the decay channels, coupling of $\chi$ to SM dimension $3$ operators
such as $Q\bar Q$, which has $\Delta B =0$, cannot provide  baryon asymmetry.
Similarly, for higher dimensional operators such as  $Q h q$, where $Q$ ($q$)
is the right (left) handed SM quark  and  $h$ is the Higgs doublet.  A
dimension $5$ operator cannot be constructed at a quark level because of the
color symmetry.  The lowest order turns out to be  the dimension $6$ operator 
$Q Q Q l$,  for which $\Delta B = \Delta L \neq 0$, which can certainly
transmute any asymmetry in $\chi$ to the quark sector. 
Thus, the global charge carried by $chi$ has to be chosen such that
$Q Q Q l$ carries the opposite one. Thus, forbidding the presence of this
operator alone on the theory and making  
the coupling $\chi Q Q Q l$  the lowest possible order for $\chi$. 
Also note this  operator can  
mediates proton decay too, unless one ensures that
$\chi$ does not  develop a vacuum expectation value, which is a severe
constraint on model building, but not a difficult one to realize. 
In the same spirit one may check those operators
which induce $n-\bar n$ oscillations. Again, an effective $\Delta B=2$ 
operator $UDDUDD$, or, such as $(QQQh)^2$ cannot be induced by, say,
integrating out $\chi$. To avoid the decay of $\chi$ into Higgs fields we
take  $h$ to be chargeless under $U(1)_\chi$.

\vskip7pt

$(4)$ {\it Can AD field be either $\phi$, or, $N$ ? }

Given the constraints, neither $N$ nor the inflaton field $\phi$ can act as 
an AD field. The auxiliary field $N$ develops a vacuum expectation value 
after the end of inflation, which would immediately induce proton decay. 
Regarding $\phi$, the bad news is that; since the asymmetry in AD field 
depends on the initial configuration, a large amplitude oscillations 
in $\phi$ can induce an undesirably large $\phi -\bar \phi$  asymmetry. 
Hence we invoke a separate AD field, a fundamental scalar field 
$\chi$ with a global charge $U(1)_{\chi}$.

\vskip7pt

We remind the readers that the inflaton energy density 
$\rho_{\rm I}$ should govern the evolution of the Universe. 
Eventually the decay products of the inflaton should
be responsible for reheating the Universe. The inflaton decays before $\chi$ 
decays via baryon violating interaction and generates
a baryon asymmetry given by
 \begin{eqnarray}
 \label{ratio}
 \frac{n_{b}}{s} \approx \frac{n_{b}}{n_{\chi}}\frac{T_{\rm r}}
 {m_{\chi}}\frac{\rho_{\chi}}{\rho_{\rm I}}\,.
 \end{eqnarray}
The final entropy released by the inflaton decay is given
by $s \approx \rho_{\rm I}/T_{\rm r}$.  The ratio
$n_{b}/n_{\chi}$ depends on the total phase accumulated
by the AD field during its helical motion in the background of an oscillating
inflaton field. In our calculation we shall always approximate 
the total phase $\sim {\cal O}(1)$. 

\vskip7pt

In theories with extra dimensions there are two choices for the AD field. It
could  either be a brane field or a bulk field like $\phi$ and $N$.  If we
assume that the AD field is a brane field, then it cannot have an effective
mass higher than the fundamental scale, and the effective initial field
amplitude  $|\chi| \leq M_{\ast}$. These constraints suggest that the energy
stored  in $\chi$  can at most be $\rho_{\chi} \approx m^2_{\chi}M_{\ast}^2$,
where  $m_{\chi} \leq M_{\ast}$. The energy density stored in $\phi$ and $N$ 
is quite  large because these fields can have initial amplitudes close to
Planck scale so that $\rho_{\rm I}\approx M_{\ast}^2M_{\rm p}^2$.  Thus we
would find
\begin{equation}
\frac{n_{b}}{s}~ \sim~\frac{T_{\rm r}}{M_{\rm p}}~
\frac{m_{\chi}}{M_{\rm p}}\quad \approx
\quad 10^{-34}~\frac{m_{\chi}}{M_{\ast}}~\ll~10^{-10}\,.
\end{equation}
for $T_{\rm r} \sim {\cal O}(1-10)$ MeV, we conclude that  $\chi$ cannot be a
brane field.

\vskip7pt

Let us therefore promote AD field to the bulk whence the energy density stored
in the AD field is $\rho_{\chi} \sim m^2_{\chi}M_{\rm p}^2$ since now $\chi$
can have large initial amplitude $|\chi| \geq M_{\ast}$. This  implies that the
{\it maximum}  baryon to entropy ratio is given by
\begin{eqnarray}
\label{final00}
\frac{n_{b}}{s} \approx \left(\frac{T_{\rm r}}{M_{\ast}}\right)
\left(\frac{m_{\chi}}{M_{\ast}}\right)\, \sim
10^{-10}\left(\frac{m_{\chi}}{{\rm 1 GeV}}\right) ,
\end{eqnarray}
where we assumed  $T_{\rm r} \sim 10$ MeV and
$M_{\ast} \sim 10$ TeV for concreteness. Thus we can achieve an adequate
baryon to entropy ratio from $\chi$ decaying into SM quarks and leptons. 
Of course, the numeric value of the ratio depends on  the initial amplitude
for $\chi$ and its mass. Note, if we set $m_\chi\sim M_{\ast}$,
we obtain the initial amplitude $\chi\sim M_{GUT}\sim 10^{16}$ GeV.

\vskip7pt

Now we can also estimate the
decay rate for $\chi$ field and $\bar \chi$ fields. The interaction 
is given by
\begin{equation}
\label{interaction}
\frac{\kappa}{M_{\ast}^2M_{\rm p}} \chi Q Q Q l\,.
\end{equation}
We sum over all possible channels; such as various color and family index
combinations which can be of the order of thousand.
On the other hand we strictly assume that inflaton is decaying 
into Higgses. Hence we estimate the decay rate as
 \be
\label{dec}
 \Gamma_\chi \approx  \left({\kappa\over f}\right)^2
 \left({m_\chi\over M_{\ast}}\right)^7~\Gamma_{\phi, N \rightarrow hh} ~.
 \ee
In order that the decay products of $\chi$ thermalizes before Nucleosynthesis
we have to assume $m_{\chi} \approx M_{\ast}$. This does not leave much freedom
for the masses which makes the model more predictive but  also demands some
level of fine tuning  if AD baryogenesis is to be successful, as we shall
discuss below.

\section{The model}

Let us now describe our model. We assume that the inflaton sector is
responsible  for breaking $U(1)_{\chi}$ charge dynamically. The AD  potential
can be written  as
\begin{eqnarray}
\label{adpot}
V_{\rm AD}(\phi,N,\chi_1,\chi_2) =
\kappa_1^2 \left(\frac{M_{\ast}}{M_{\rm p}}\right)^2 N^2(\chi_1^2+\chi_2^2)
 +\frac{\kappa_2^2}{4}\left(\frac{M_{\ast}}{M_{\rm p}}\right)^2
 (\chi_1^2+\chi_2^2)^2\, \nonumber \\
+\kappa_3 ^2\left(\frac{M_{\ast}}{M_{\rm p}}\right)^2\phi N 
(\chi_1^2-\chi_2^2)\,,
\end{eqnarray}
where $\kappa_1,\kappa_2,\kappa_3$ are constants, and $\chi_1$ and $\chi_2$ are
the real and imaginary components of the complex field $\chi$. 
Note, all the terms are
Planck mass  suppressed, because Eq.~(\ref{adpot}) depicts an effective 
four dimensional potential derived from higher dimensional Lagrangian
by integrating out the extra spatial dimensions. 
From Eq.~(\ref{adpot}), it is evident that since during
inflation the auxiliary field is trapped in the false vacuum $N=0$, this
renders $\chi_1$ and $\chi_2$ massless and the AD potential becomes almost
flat. We also notice that the last term is also not present during inflation.
Therefore, the only contributions comes from the self coupling which obviously
allows  the AD field to evolve. The details of the field evolution shall be
discussed in the  coming subsections. Note that we do not include a term of the
type $(M_{\ast}/M_{\rm p})^2 \phi^2 |\chi|^2$ as it would ruin both inflation
and baryogenesis because of the large-amplitude oscillation of $\chi$; this is
a fine tuning we have to perform on our model for which we have no dynamical
explanation.


\subsection{Constraining the initial amplitude for the AD field}

Dynamical evolution of multi scalar fields induces density perturbations of
two kinds; adiabatic and isocurvature. Since, in our case $\chi_1$ and
$\chi_2$ are massless during inflation they contribute to the adiabatic
fluctuations. In order to obtain a scale invariant spectrum, the fluctuations
should be mainly  generated by $\phi$ and $N$. However, this imposes an upper
bound on the allowed amplitude for the AD field at the time when the modes
relevant for large scale  structure formation are leaving the horizon during
inflation, which happens in our case  during the last $43$ e-foldings of
inflation. 

\vskip7pt

In the regime when
the main inflationary potential is due to the inflaton sector, the 
slow-roll parameter $|\eta|$ can be written as \cite{kari} 
\begin{equation}
\label{dev}
|\eta| \approx \frac{M_{\rm p}^2}{8\pi}
 \frac{V^{\prime \prime}(\phi,N)}{V(\phi,N)}
-\frac{M_{\rm p}^2}{8\pi}\frac{V_{\rm AD}^{\prime}V_{\rm AD}^{\prime \prime}}
{V(\phi,N)V^{\prime}(\phi,N)}\,,
\end{equation}
where prime in $V_{\rm AD}$ denotes derivative with respect to AD field
$\chi$. According to Eq.~(\ref{con}), the second term in the above equation
must be smaller than $10^{-10}$. This leads to    
\begin{equation}
\label{init}
|\chi(0)|^{5} \leq 32\pi\times 10^{-10}\times 
 \frac{\lambda^2}{\kappa_2^4}
 \left(\frac{N_0}{M_{\ast}}\right)^5 M_{\rm p}^5\,.
\end{equation}
If $\lambda, \kappa_2 \sim {\cal O}(1)$, and $N_0\sim M_{\ast} $, we obtain an
initial amplitude for $|\chi(0)| \leq 10^{-2}M_{\rm p}$. This is a constraint
which we should bear in mind while estimating the total baryon asymmetry. Note,
for a reasonably fast phase transition $N_0 < M_{\ast}$, the amplitude for
$\chi(0)$ becomes even smaller than the above limit.

\vskip7pt

Similarly one might expect isocurvature perturbations arising from the AD
field. Notice that the angular direction of AD field is effectively massless,
but gains a dynamical mass $\geq H$ just after inflation. This is unlike the
case of a supersymmetric AD baryogenesis where the field remains massless even
after the end of inflation provided there is no supergravity correction to the
$U(1)$ violating term. Nevertheless, in our case we would expect isocurvature
fluctuations during inflation because there are more than one dynamical scalar
fields present during inflationary phase. It is well known that in the case of
isocurvature fluctuations it is easier to constrain the spectral index
\cite{isoliddle}, rather than the amplitude of perturbations. This restricts
the initial amplitude for the AD field very much in a similar vein as in
Eq.~(\ref{init}). However, since we are assuming  that the main contribution to
the density perturbations is coming from the adiabatic sector, rather than the
isocurvature fluctuations, we would not expect any further improvement on the
limit we have already obtained upon the initial amplitude for  
$\chi_1$ and $\chi_2$ from Eq.~(\ref{init}).

\subsection{Dynamics of AD field}

The dynamics of $\chi$  is  complicated. Even though $\chi$ has no effective
mass during inflation, it has an effective quartic self coupling which
determines  its dynamics  (see Eq.~(\ref{adpot})).  The equations of motion in
terms of the component fields $\chi_1,\chi_2$ are given by
\begin{eqnarray}
\label{dyn1}
\ddot \chi_1+3H\dot\chi_1 &=&
 -\kappa_2^2\left(\frac{M_{\ast}}{M_{\rm p}}\right)^2
 (\chi_1^2+\chi_2^2)\chi_1 \,, \\
\label{dyn2}
\ddot \chi_2+3H\dot\chi_2 &=&
 -\kappa_2^2 \left(\frac{M_{\ast}}{M_{\rm p}}\right)^2
(\chi_1^2+\chi_2^2)\chi_2 \,.
\end{eqnarray}
We remind that $\chi_1$ and $\chi_2$ are the components of a 
a bulk field, therefore, they would have naturally taken initial values close
to the fundamental scale in higher dimensions but close to the Planck scale in
four dimensions. From the point  of view of four dimensions, the fields simply
roll down from the Planck scale because the self coupling induces a curvature
terms for $\chi_1$ and $\chi_2$ fields.  We note that the suppression 
in the couplings is very small $\sim (M_{\ast}/M_{\rm p})^2$  if  
$\kappa_2 \sim {\cal O}(1)$. The fields are very weakly self coupled.  
However, in the process of rolling down the potential their effective 
running mass becomes of $\sim H$,  given by Eq.~(\ref{hub0}). When this 
happens their dynamics is effectively frozen at a particular amplitude 
which observes the constraint derived independently in the
earlier section, see Eq.~(\ref{init}).

\vskip7pt

As we have discussed, we may assume that $\chi_1$ and $\chi_2$ collectively
follow a similar trajectory. The dynamics of $|\chi|$ freezes when
$3H|\dot\chi| \approx -\kappa_2^2(M_{\ast}/M_{\rm p})^2|\chi|^3$. 
This immediately gives us  a simple solution for the largest 
values of the fields:
\begin{equation}
\label{val}
\chi_1(0) \sim \chi_2(0) \leq 
\frac{1}{\kappa_2}\left(\frac{M_{\rm p}}{M_{\ast}}
\right)\sqrt{\frac{3H}{2\Delta t}}
\approx \frac{\lambda}{\kappa_2}\sqrt{\frac{4\pi}{N_{\rm e}}}M_{\rm p}\,,
\end{equation}
where $N_{\rm e}=H\Delta t$ is the total number of e-foldings, which could be
larger  than $\sim 130$. The total number also takes into account of inflation
during the radion stabilization, but for the structure formation it is the
last $43$ e-foldings of inflation which matters. For a simple estimate, if 
$\lambda \sim \kappa_2$, we obtain $|\chi(0)|\leq 10^{18}$ GeV, in well
accordance with Eq.~(\ref{init}).  When inflation comes to an end all the
fields begin oscillations; for $\phi$ and $N$, the initial amplitudes are given
by Eq.~(\ref{amp0}), while for $\chi_1$ and $\chi_2$ the initial amplitudes
are determined by Eq.~(\ref{init}).

\vskip7pt

The post-inflationary dynamics of $\chi_1$ and $\chi_2$ depends on $\phi$ and
$N$, which are both oscillating. As a result $\chi_1$ and $\chi_2$ become
massive fields (see Eq.~(\ref{adpot})). This leads to the equations of motion
\begin{eqnarray}
\label{dyn3}
\ddot \chi_1+3H\dot\chi_1 &=&
 -2\kappa_1^2\left(\frac{M_{\ast}}{M_{\rm p}}\right)^2 N^2\chi_1-
\kappa_2^2\left(\frac{M_{\ast}}{M_{\rm p}}\right)^2(\chi_1^2+\chi_2^2)\chi_1
-2\kappa_3^2 \left(\frac{M_{\ast}}{M_{\rm p}}\right)^2 \phi N \chi_1\,, \\
\label{dyn4}
\ddot \chi_2+3H\dot\chi_2 &=&
  -2\kappa_1^2\left(\frac{M_{\ast}}{M_{\rm p}}\right)^2 N^2\chi_2 - 
\kappa_2^2\left(\frac{M_{\ast}}{M_{\rm p}}\right)^2(\chi_1^2+\chi_2^2)\chi_2
+2\kappa_3^2 \left(\frac{M_{\ast}}{M_{\rm p}}\right)^2\phi N\chi_2 \,.
\end{eqnarray}
There is an effective mass term for $\chi$ fields which is
again field dependent. The last term in the above equations comes
with an opposite sign and this is  responsible
for giving rise to a small asymmetry in $\chi$ over $\bar\chi$.  As a
simplest approximation, we neglect the back reaction of $\chi$ fields on 
the background fields, which is actually true as long as $N, \phi$ are 
responsible for generating the final entropy of the Universe. This 
assumption simplifies the situation and allows us to estimate 
$\chi$ asymmetry created by the motion of $\chi_1$ and $\chi_2$ analytically.

\vskip7pt

We have earlier pointed out that $\phi$ and $N$  oscillate several times
before they decay at the reheating temperature. While the
initial oscillations are quite large $\sim M_{\rm p}$, the energy
density stored in the oscillations is quite small due to the presence of 
very small couplings of order $(M_{\ast}/M_{\rm p})^2 \sim 10^{-28}$. 
Likewise, $\chi_1$ and $\chi_2$ also oscillate with a large amplitude and 
they also go through several oscillation periods before they decay
with a rate  given by Eq.~(\ref{dec}). By substituting Eq.~(\ref{amp0}) 
in Eq.~(\ref{adpot}), we may estimate the effective masses
\begin{equation}
\label{ADmass}
m_{\chi_1} \approx \sqrt{2}\kappa_1 N_{0} \,, \quad \quad
m_{\chi_2} \approx \sqrt{2}\kappa_1 N_{0}\,.
\end{equation}
The coupling constants $\kappa_1, \kappa_2$ can be tuned to obtain  
$m_{\chi_1} \sim m_{\chi_2} \approx M_{\ast}$ 
in order to match the decay rate of
$\Gamma_{\chi} \sim \Gamma_{\phi}\sim \Gamma_{N}$. 
This finally ensures that
the  AD field decays along with $\phi$ and $N$ fields, and this
also prevents the decay of $\phi$ and $N$ into $\chi$ fields. 
If we compare Eqs.~(\ref{amp1}),  and (\ref{ADmass}), we notice that 
all the masses are of same order with a small 
variation due to different couplings.
If we set  $\kappa_1 \geq \kappa_2 \sim \lambda \sim g$, we can ensure that
$\chi_1$ and $\chi_2$ go through more oscillation periods than the background
fields $\phi$ and $N$. Here we tacitly assume that the oscillations in $\chi$ 
do not have any dynamical back reaction on $\phi$ and $N$ oscillations. This is
true if  the field couplings are weak and indeed, in our case the couplings
such as $\kappa_1^2(M_{\ast}/M_{\rm p})^2$, $\kappa_3^2(M_{\ast}/M_{\rm p})^2$ 
are extremely small even if we choose 
$\kappa_1 \sim \kappa_3 \approx {\cal O}(1)$.
Such a weak coupling also ensures that the field trajectories do not give rise
to a chaotic behavior. In order to avoid any chaotic behavior it is
necessary that all the fields oscillate with a similar frequency around the 
bottom of their respective potential. Hence we choose 
$\kappa_1 \equiv \lambda \equiv g$ while $\kappa_2$ is constrained by the
expressions Eqs.~(\ref{init}), and (\ref{val}), and $\kappa_3$ shall be
constrained by the baryon to entropy ratio. However, in order to have a helical
motion for $\chi_1$ and $\chi_2$, we ought to have  $\kappa_1 >\kappa_3$, and
$\kappa_2 > \kappa_3$.

\vskip7pt

We still have to estimate the classical behavior of $\chi_1$ and $\chi_2$,
which do not develop any vacuum expectation value. Their classical evolution
can be  estimated by
\begin{equation}
\label{adev}
\chi_1(t) \sim \chi_1(0) A(t)\cos(m_{\chi_1}t)\,, \quad \quad 
\chi_2(t) \sim \chi_2(0) A(t)\cos(m_{\chi_2}t)\,,
\end{equation}
where the  amplitudes $\chi_1(0)$ and $\chi_2(0)$ 
are constrained by Eqs.~(\ref{init}) and (\ref{val}).
Both the amplitudes decrease in time as $A(t)\propto 1/t$.
The frequency of the oscillations are determined by Eq.~(\ref{ADmass}).  In
what follows we shall always assume that the relative phase between the  fields
$\chi_1$ and $\chi_2$ is of order ${\cal O}(1)$. We now have all the tools  to
address the generation of $\chi \sim \bar \chi$ charge asymmetry from the
classical evolution of all the fields. This shall provide us with a final
baryon to entropy  ratio in this particular model.


\subsection{Baryon to entropy ratio}

The $CP$ violation in our model is given by the last term in
Eq.~(\ref{adpot}).  For a charged scalar field this is equivalent to $C$
violation. The $B$ violation  arise via the decay of $\chi$, because the decay
products have $\Delta B \neq 0$,  and we have a non-trivial helical
oscillations in $\chi$ which accumulates net $CP$ phase which is transformed
into asymmetric $\chi$. The net $\chi$ asymmetry; $n_{\chi}$ can be calculated
by evaluating the Boltzmann equation: 
\begin{eqnarray}
n_{\chi}= \frac{i}{2}\left(\dot \chi^{\ast}\chi-
\chi^{\ast}\dot \chi\right)\,.
\end{eqnarray}
With the help of Eqs.~(\ref{dyn3}) and (\ref{dyn4}), we can rewrite the above
expression as
\begin{eqnarray}
\label{lep}
\dot n_{\chi}+3Hn_{\chi}=4\kappa_3^2 \left(\frac{M_{\ast}}{M_{\rm p}}\right)^2
\langle N(t)\phi(t)\rangle \chi_1(t)\chi_2(t)\,.
\end{eqnarray}
The right hand side of the above equation is the source term which generates a
net $\chi$ asymmetry through a non-trivial motion of $\chi_1$ and
$\chi_2$ fields. We integrate Eq.~(\ref{lep}) from $t_0$ which corresponds
to the end of inflation up to a finite time interval.
\begin{eqnarray}
\label{integral}
n_{\chi} a^3 &=& 4\kappa_3^2 \left(\frac{M_{\ast}}{M_{\rm p}}\right)^2
\int_{t_0}^{t}\langle N(t^{\prime}) \phi(t^{\prime})\rangle a^3(t^{\prime})
\chi_{1}(t^{\prime})\chi_{2}(t^{\prime})dt^{\prime}\,.
\end{eqnarray}
The upper limit of integration signifies the end of reheating. We assume that
the oscillations continue until the fields decay completely. Before we perform
the integration, we notice that the integrand decreases in time.  This can be
seen as follows ; first of all notice that the approximate number of
oscillations are quite large before the fields decay 
$\sim (M_{\rm p}/M_{\ast})^2$. This allows us to average $\phi$ and $N$
oscillations; from Eqs.~(\ref{ev0}) and (\ref{ev2}) we get the time dependence
$\langle \phi N \rangle \sim \Phi^2(t) \sim 1/t^2$.  Similarly, from
Eq.~(\ref{adev}),  $\langle \chi_1 \chi_2 \rangle \sim 1/t^2$ provides 
another suppression. While taking care of the expansion, where the scale factor
behaves like $a(t)\sim t^{2/3}$, the overall
behavior of the integrand follows $\sim 1/t^2$. 
This suggests that the maximum contribution to $\chi$ asymmetry comes only at
the initial times when $t_0 \sim 1/H_0$, where $H_0$ is determined by
Eq.~(\ref{hub0}). The right hand side of the above equation turns out to be
\begin{equation}
\label{integral1}
n_{\chi} \approx \frac{2}{27}\kappa_3^2 \frac{M_{\ast}^2 |\chi(0)|^2}{H_0}\,.
\end{equation}
We have assumed that the total $CP$ phase, which is given by two factors: an
initial phase determined arbitrarily during inflation and the final dynamical
phase which is accumulated during the oscillations, is of the order $\sim {\cal
O}(1)$. We have also assumed $N_0 \sim M_{\ast}$.

\vskip7pt

The final ratio of $\chi$ number density produced and the entropy is given by 
\begin{eqnarray}
\label{final}
\frac{n_{\chi}}{s} \approx 
{2\kappa_3^2\over 27\lambda^2} \left({|\chi(0)|\over M_{\rm p}}\right)^2
 \left({T_{\rm r}\over H_0}\right) \leq
\frac{2\sqrt{6\pi}}{27}\left(\frac{\kappa_3}{\kappa_2}
\right)^2\left(\frac{1}{\lambda N_{\rm e}}\right) 
\left(\frac{T_{\rm r}}{M_{\ast}}\right)\, ,
\end{eqnarray}
where we have used the fact that $s\propto a^{3}$ in our case.
This is the final expression for $\chi$ asymmetry produced during the helical 
oscillations of $\chi$ and is to be compared with the observed baryon asymmetry,
whose range is 
$4(3)\times 10^{-10} \leq \eta_B\equiv n_B/n_\gamma 
\leq 7(10)\times 10^{-10}$~\cite{ratio}.
The upper bound on the baryon to entropy ratio given by 
Eq.~(\ref{final}) depends on the amplitude of $\chi_1$ and $\chi_2$,
which were frozen during inflation, it also depends on the 
number of e-foldings, which could be $\sim 100$. In order to evaluate 
Eq.~(\ref{final}),  we may take an example: $T_{\rm r} \sim 100$~MeV, 
and $M_{\ast}\sim 100$~TeV. This gives an   
asymmetry of order $\sim 10^{-10}$, if we choose the couplings of order one.
The final asymmetry in $\chi$ is transferred  to the SM quarks via baryon
number violating interaction mentioned in  Eq.~(\ref{interaction}). The
asymmetry is injected into thermal bath along with the  inflaton decay
products. It is essential that the thermalization takes
place after AD field has decayed in order not to wash away the total
baryon asymmetry. 

\vskip7pt

Finally, we mention that our model is quite generic.
The robustness of the model is that it can work for arbitrary number of extra
spatial dimensions. The model predicts baryogenesis just above nucleosynthesis
scale. The model does not rely on first order phase transition of
electro weak baryogenesis, nor does it depend on sphaleron transition rate.
Our approach is similar to Ref.~\cite{zurab}, where this model has been
embedded in a supersymmetric theory in four dimensions. Like in the case of AD
baryogenesis~\cite{kari}, our setup can be tested by the forthcoming
accurate measurements of the spectral index of the microwave background; our
model predicts a very flat spectrum.


\section{conclusion}

In summary, we have presented a natural mechanism of baryogenesis in the
context of low quantum gravity scale with large compact extra dimensions. Our
mechanism is generic and it is independent of the fundamental scale and the
number of compact extra dimensions.  This mechanism does not rely on any extra
assumption other than invoking a fundamental scalar field that lives in the 
$4+d$ dimensional space time.  The baryogenesis scheme can work at any
temperature lower than the electro weak scale. This is because of the presence
of non-renormalizable couplings which automatically reheats the Universe with a
temperature close to BBN. This feature is independent of the number of extra
dimensions. Our mechanism does not depend on the rate of sphaleron transition
and also does not rely on leptogenesis. 

\vskip7pt

The three requirements for baryogenesis is fulfilled in our case as follows.
The $C$ and $CP$ violation is a dynamical process in our case. The AD field has
a charge associated with a global $U(1)_{\chi}$ symmetry. This symmetry is 
broken once the AD field gains a dynamical mass from its coupling to the 
inflaton sector. This happens once inflation comes to an end and when all the
fields start oscillating around their respective minima. The dynamics of the AD
field is solely responsible for $CP$ violation and the initial $CP$ phase is
determined during inflation which we have assumed to be of order one. During 
post inflationary era the real and the imaginary parts of the AD field has 
different equations of motion and as a result the motion of $AD$ field is
helical in the field space. During this helical motion the AD field accumulates
a net asymmetry in $\chi$ over $\bar \chi$. This asymmetry is then transferred
to the SM sector via a  baryon violating interaction which does not allow fast
proton decay.

\vskip7pt

The decay of inflaton into Higgses and their subsequent decay generates a net
entropy in the Universe. The thermalization of the Universe is quite late and
the final reheat temperature is as low as $100$ MeV for a fundamental scale as
small as 100 TeV. The AD field decays into quarks and leptons directly
imparting the total baryon asymmetry it has  generated during the helical
motion, obviously, the net $CP$ phase generated during  this helical motion is
understood to be also of order one. We note that the baryon to entropy ratio is
determined by the ratio of the reheat temperature and the  fundamental scale.
The ratio also depends on the number of e-foldings. This  tells us that the
evolution of the massless AD field during the inflationary stage also counts in
producing a net baryon asymmetry, through fixing the initial amplitude of the
AD field. The most important aspect of our model is that the model parameters
are very constrained from the density  perturbations during inflation. The
initial amplitude of the AD field is also constrained in order not to depart
from the scale invariance of density  perturbations measured by the satellite
and balloon experiments. The
constraint which we apply on the amplitude of  the AD field comes purely from
adiabatic perturbations. However, presence of many fields automatically
introduces a possibility of isocurvature fluctuations which we shall work out
in future. In this sense, our model can be falsifiable.
As a final note we mention that our dynamical 
mechanism of producing baryon asymmetry can be used even in four dimensions and
the mechanism is capable of generating baryon asymmetry at a scale
much lower than the electro weak scale.


\section{acknowledgements}

The authors are thankful to Mar Bastero-Gil, Zurab Berezhiani, Pierre Binetruy,
Sacha Davidon,  Alexandre Dolgov, and Katrin Heitmann for helpful discussions.
R.A. is supported by ``Sonder-forchschungsberich 375 f$\ddot{\rm u}$r
Astro-Teilchenphysik'' der Deutschen Forschungsgemeinschaft, K.E. partly by the
Academy of Finland  under  the contract 101-35224, and A.M. acknowledges the
support of {\it The Early Universe network} HPRN-CT-2000-00152. A.M.
acknowledges the  hospitality of the Helsinki Institute of Physics where part
of the work has been  carried out.


\end{document}